\documentclass[aps,prx,superscriptaddress,twocolumn,showpacs, longbibliography]{revtex4-2}

\usepackage{graphicx}
\usepackage{dcolumn}
\usepackage{bm}
\usepackage{amsmath}
\usepackage{siunitx}
\usepackage{longtable}

\begin{document}

\title{Evolution of magnetic-field-induced ordering in the layered structure quantum Heisenberg  triangular-lattice antiferromagnet Ba$_3$CoSb$_2$O$_9$}

\author{N.~A.~Fortune}
\email{nfortune@smith.edu}
\affiliation{Department of Physics, Smith College, Northampton, Massachusetts 01063, USA}

\author{Q. Huang}
\affiliation{Department of Physics and Astronomy, University of Tennessee, Knoxville, Tennessee 37996-1200, USA}

\author{T. Hong}
\affiliation{Neutron Scattering Division, Oak Ridge National Laboratory, Oak Ridge, TN 37831, USA}

\author{J. Ma} 
\affiliation{Laboratory of Artificial Structures and Quantum Control, School of Physics and Astronomy, Shanghai Jiao Tong University, Shanghai 200240, China}
\affiliation{Shenyang National Laboratory for Materials Science, Institute of Metal Research, Chinese Academy of Sciences, 110016 Shenyang, China}

\author{E.~S.~Choi}
\affiliation{National High Magnetic Field Laboratory, Florida State University, Tallahassee, FL 32310-3706, USA}

\author{S.~T.~Hannahs}
\affiliation{National High Magnetic Field Laboratory, Florida State University, Tallahassee, FL 32310-3706, USA}

\author{Z. Y. Zhao}
\affiliation{State Key Laboratory of Structure Chemistry, Fujian Institute of Research on the Structure of Matter, Chinese Academy of Science, Fuzhou, Fujian 350002, People's Republic of China}

\author{X. F. Sun}
\email{xfsun@ustc.edu.cn}
\affiliation{Department of Physics, Hefei National Laboratory for Physical Sciences at Microscale, and Key Laboratory of Strongly-Coupled Quantum Matter Physics (CAS), University of Science and Technology of China, Hefei, Anhui 230026, People's Republic of China}
\affiliation{Institute of Physical Science and Information Technology, Anhui University, Hefei, Anhui 230601, People's Republic of China}

\author{Y.~Takano}
\email{takano@phys.ufl.edu}
\affiliation{Department of Physics, University of Florida, Gainesville, Florida 32611-8440, USA}

\author{H. D. Zhou}
\email{hzhou10@utk.edu}
\affiliation{Department of Physics and Astronomy, University of Tennessee, Knoxville, Tennessee 37996-1200, USA}
\affiliation{National High Magnetic Field Laboratory, Florida State University, Tallahassee, FL 32310-3706, USA}

\date{\today}

\begin{abstract}
Quantum fluctuations in the effective spin-1/2 layered structure triangular-lattice quantum Heisenberg antiferromagnet Ba$_3$CoSb$_2$O$_9$ lift the classical degeneracy of the antiferromagnetic ground state in magnetic field, producing a series of novel spin structures for magnetic fields applied within the crystallographic $ab$ plane, including a celebrated collinear `up-up-down' spin ordering with  magnetization equal to 1/3 of the saturation magnetization over an extended field range. Theoretically unresolved, however, are the effects of interlayer antferromagnetic coupling and transverse magnetic fields on the  ground states of this system. Additional magnetic-field-induced phase transitions are theoretically expected and in some cases have been  experimentally observed, but details regarding their number, location,  and physical character appear inconsistent with the predictions of existing models. Conversely, an absence of experimental measurements as a function of magnetic-field orientation has left other key predictions of these models untested. To address these issues, we have used specific heat, neutron diffraction, thermal conductivity, and magnetic torque measurements to map out the phase diagram as a function of magnetic field intensity and  orientation relative to the crystallographic $ab$ plane. For $H||ab$, we have discovered  an additional, previously unreported  magnetic-field-induced phase transition at low temperature and an unexpected tetracritical point in the high field phase diagram, which —  coupled with the apparent second-order nature of the phase transitions — eliminates several theoretically proposed spin structures for the high field phases.  Our calorimetric measurements as a function of magnetic field orientation are in general agreement with theory for field-orientation angles close to plane parallel ($H||a$) but diverge at angles near plane perpendicular; a predicted convergence of two phase boundaries at finite angle and a corresponding change in the order of the field induced phase transition is not observed experimentally. Our results emphasize the role of  interlayer coupling in selecting and stabilizing field-induced phases, provide new guidance into the nature of the magnetic order in each phase,  and reveal the need for new physics to account for the nature of magnetic ordering in this archetypal 2D spin-1/2 triangular lattice quantum Heisenberg antiferromagnet.    \end{abstract}

\maketitle

\section{Introduction}

The layered structure transition-metal oxide $\text{Ba}_3\text{Co}\text{Sb}_2\text{O}_9$  is a nearly ideal realization of an isotropic two-dimensional (2D) spin-$\frac{1}{2}$ triangular-lattice quantum Heisenberg antiferromagnet (TLHAF).  The 2D triangular arrangement of nearest neighbor $\text{Co}^{2+}$ ions with effective spin 1/2 frustrates the otherwise expected classical magnetic ordering of the spins in an applied magnetic field and its inversion symmetry means that there is no competing Dzyaloshinkii-Moriya interaction between nearest neighbor $\text{Co}^{2+}$ ions. 
As a  result, the magnetic ordering in $\text{Ba}_3\text{Co}\text{Sb}_2\text{O}_9$  arises from zero-point motion that  lifts the degeneracy of the classical ground state \cite{chubukov1991quantum}. Interlayer coupling leads to long-range order at a finite temperature at zero field \cite{doi2004structural} but also alters the nature of magnetic ordering in an applied field. For this reason, $\text{Ba}_3\text{Co}\text{Sb}_2\text{O}_9$  serves as an experimental touchstone for  quantum mechanical models of antiferromagnetic ordering in 2D materials with weak interlayer  exchange \cite{collins1997triangular, starykh2015unusual}. 

Of particular interest in $\text{Ba}_3\text{Co}\text{Sb}_2\text{O}_9$ are three questions: (1) how phase transitions between different magnetic spin structures are induced by an applied magnetic field, (2) what type of spin structure exists in each magnetic-field-induced phase, and (3) how the phases and spin structures vary  with  magnetic field orientation.

\begin{figure}[hb]
\centering\includegraphics[clip,width=8cm]{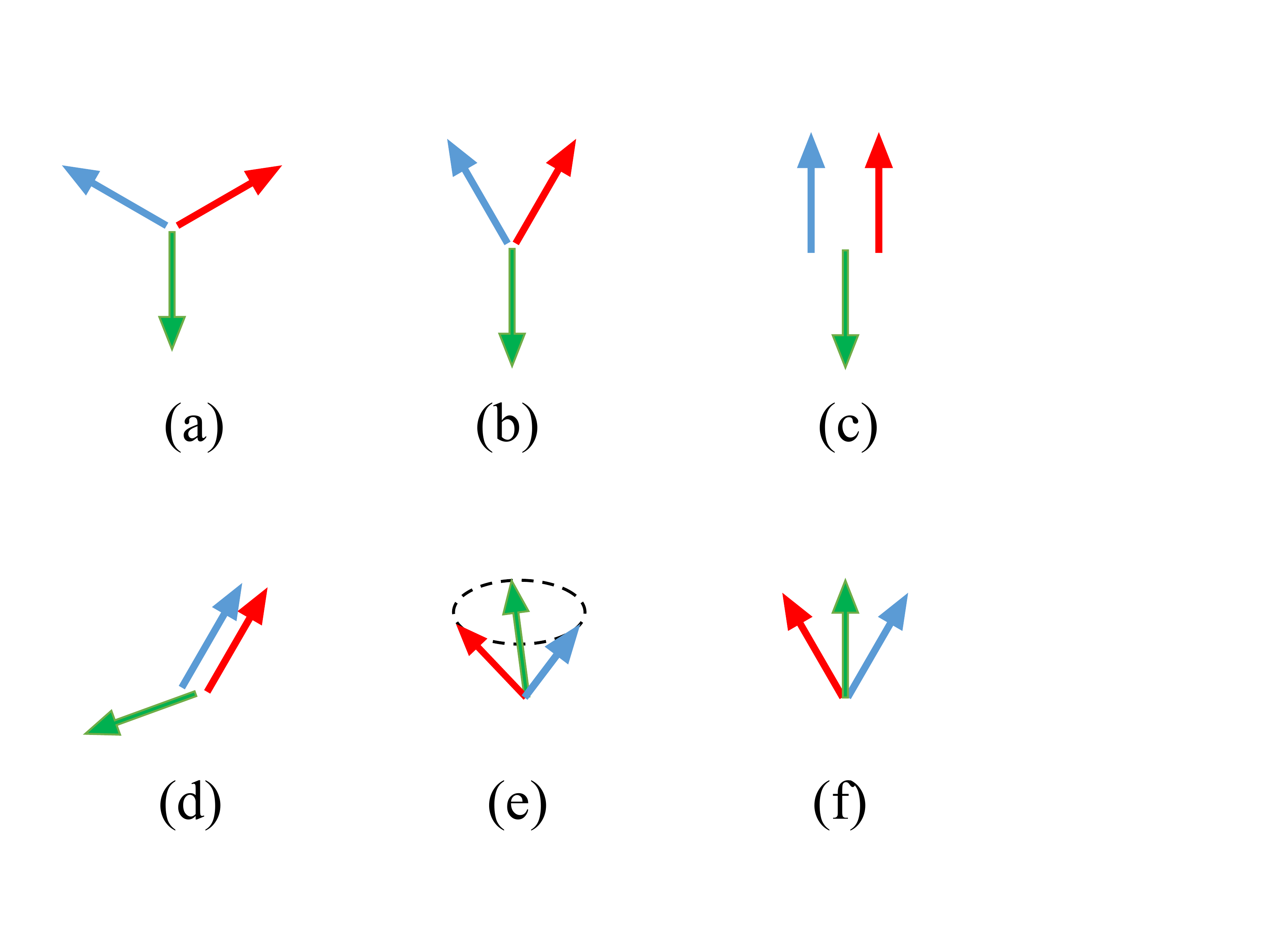}
\caption{(Color online) Theoretically proposed spin orderings for the quantum $S = 1/2$ Heisenberg triangular antiferromagnet in the isotropic 2D limit. Structures (a - d) are the coplanar `Y', `UUD', and `V' co-linear spin ordering favored by quantum fluctuations, while (e) represents the competing classically favored non-coplanar 3D cone (umbrella) spin arrangement. Structure (f) represents the alternative high field coplanar  spin ordering $\Psi$ (also referred to as a `fan' structure).   In zero field, theory predicts that the spins order in a \ang{120} arrangement  within the easy plane ; phase transitions from the \textit{Y} (b) to the up-up-down \textit{UUD}  (c) to the \textit{V} spin structures (d) occur as the  magnetic field is increased \cite{chubukov1991quantum}. 
}
\label{fig:2D_spin}
\end{figure}

In the zero-temperature, isotropic 2D  limit,  $S=1/2$ spins in a frustrated triangular lattice  arrange at zero field in the three-fold degenerate \ang{120} coplanar spin configuration shown in Fig.~\ref{fig:2D_spin} a; in the case of the easy-plane antiferromagnet $\text{Ba}_3\text{Co}\text{Sb}_2\text{O}_9$ studied here, the spins lie in the easy ($ab$) plane, normal to the $c$ axis. 

In an externally applied magnetic field, theory predicts this state will continuously evolve into the first of three quantum fluctuation stabilized phases: a coplanar  phase in which the spins order in the shape of a \textit{Y}, as shown in Fig.~\ref{fig:2D_spin}b.  The angle between the two spins forming the top branches of the \textit{Y} depends on the strength of the applied magnetic field \cite{chubukov1991quantum}. 

At an applied field equal to approximately 3/10 of the saturation magnetic field $H_s$,  a phase transition to the collinear up-up-down (\textit{UUD}) spin structure shown in Fig.~\ref{fig:2D_spin}c occurs, 
one signature of the \textit{UUD} phase being a plateau in the magnetization at 1/3 of the saturation magnetization.   In $\text{Ba}_3\text{Co}\text{Sb}_2\text{O}_9$ , this plateau is experimentally observed between 10 - 15 tesla for fields directed within the easy plane 
\cite{shirata2012experimental, susuki2013magnetization, sera2016s}.  

The third predicted phase is a different coplanar  spin ordering denoted \textit{V}: two of the three spins share a common orientation, thereby forming a rotated V shape, as shown in Fig.~\ref{fig:2D_spin}d. These three phases are followed by a final field induced transition at $H_s$ to the fully polarized state.

\begin{figure}[hb]
\centering\includegraphics[clip,width=8cm]{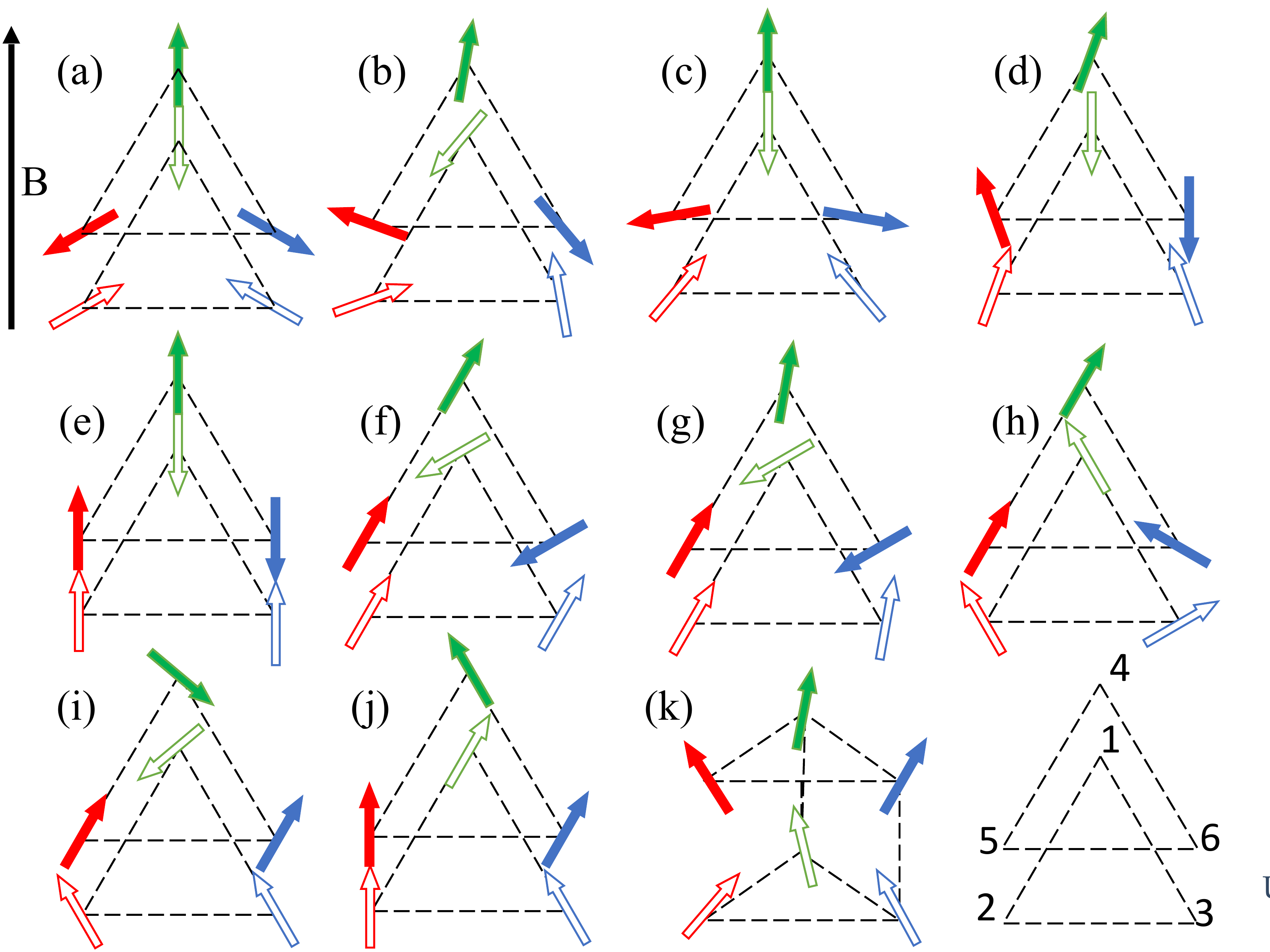}
\caption{(Color online) Theoretically proposed spin orderings for a quantum $S = 1/2$ Heisenberg triangular antiferromagnet with interlayer coupling in a 2 layer model. Spins 1 - 3 correspond to layer 1; spins 4 - 6 to layer 2. The orderings are named as follows:  (a) the zero field \ang{120} structure, (b) distorted combined Y ,(c) alternating ordered Y, (d) parallel ordered Y, (e) UUD, (f) V, (g) distorted V, (h) V$^{\prime}$, (i) staggered V$^{\prime}$, (j) $\psi$ (fan) spin structure.  Ordering (k) corresponds to the classically expected cone (umbrella) phase.} 
\label{fig:2D_spin_with_interlayer}
\end{figure}

Incorporating interlayer exchange and weak in-plane anisotropy is expected to result in additional ordered phases at high field, as the spins can now alternate directions on adjacent layers, and the phase diagram becomes magnetic-field-orientation dependent   \cite{gekht1997JETP, chen2013groundstates, starykh2014nearsaturation,yamamoto2014quantum,     koutroulakis2015quantum, yamamoto2015microscopic, liu2019microscopic}. 
A representative collection of the various spin orderings that have been proposed to occur when interlayer coupling is taken into account for magnetic fields aligned with the easy plane is shown in Fig.~\ref{fig:2D_spin_with_interlayer}.

In this article, we report results of specific heat, magnetic torque, thermal conductivity, and neutron scattering measurements on $\text{Ba}_3\text{Co}\text{Sb}_2\text{O}_9$ as a function of magnetic field intensity and field orientation within the 2D plane ($H||a$), normal to the 2D plane ($H||c$), and as a function of angle $\phi$ for rotations from $c$ ($\phi = \ang{0}$)  to $a$ ($\phi = \ang{90}$). We present evidence for additional phase transitions in both the low and high field limits below the saturation field $H_s$. We use thermodynamic and symmetry-based arguments to constrain the range of possible spin structures for the high field phase — in some cases ruling out previously assumed configurations — and suggest new physics beyond the standard $J-J_z-J'$ weakly in-layer anisotropic coupled layer model that might account for our results. 

\section{Experiment}

Single crystals of $\text{Ba}_3\text{Co}\text{Sb}_2\text{O}_9$ were grown by using the optical floating-zone technique. Specific-heat measurements $C_p(H, T, \phi)$ were carried out on 1.0 and 2.5 mg crystals as a function of temperature, magnetic field, and field orientation between 0.3 K and 6 K in fields between 0 and 35 tesla using custom-built single-axis rotation micro-calorimeters  \cite{hannahs2003SCMcal, fortune2014PDFcal} cross-calibrated in magnetic field  to correct for magnetoresistance of the thermometers \cite{fortune2000RSI}. The thermal conductivity, $\kappa$, was measured by using a ``one heater, two thermometer'' technique \cite{shen2016quantum, li2020possible}. The magnetic torque, $\tau$, was measured by using a 13-$\mu$m-thick CuBe cantilever. The single crystal neutron diffraction measurements were made using a $^3$He insert at the Cold Neutron Triple-Axis Spectrometer CG4C at the High Flux Isotope Reactor, with the  crystal  aligned on the [HHL] scattering plane and in a vertically directed magnetic field.

\subsection{Results and Discussion for $H||a$}

We begin with a discussion of our results for the magnetic field applied within the easy ($ab$) plane, parallel to the $a$ axis. For this orientation, we expect the spins to remain within the easy plane, starting from the \ang{120} easy-plane spin configuration at zero field. 

\textit{}

\begin{figure}[ht]
\centering\includegraphics[clip,width=8cm]{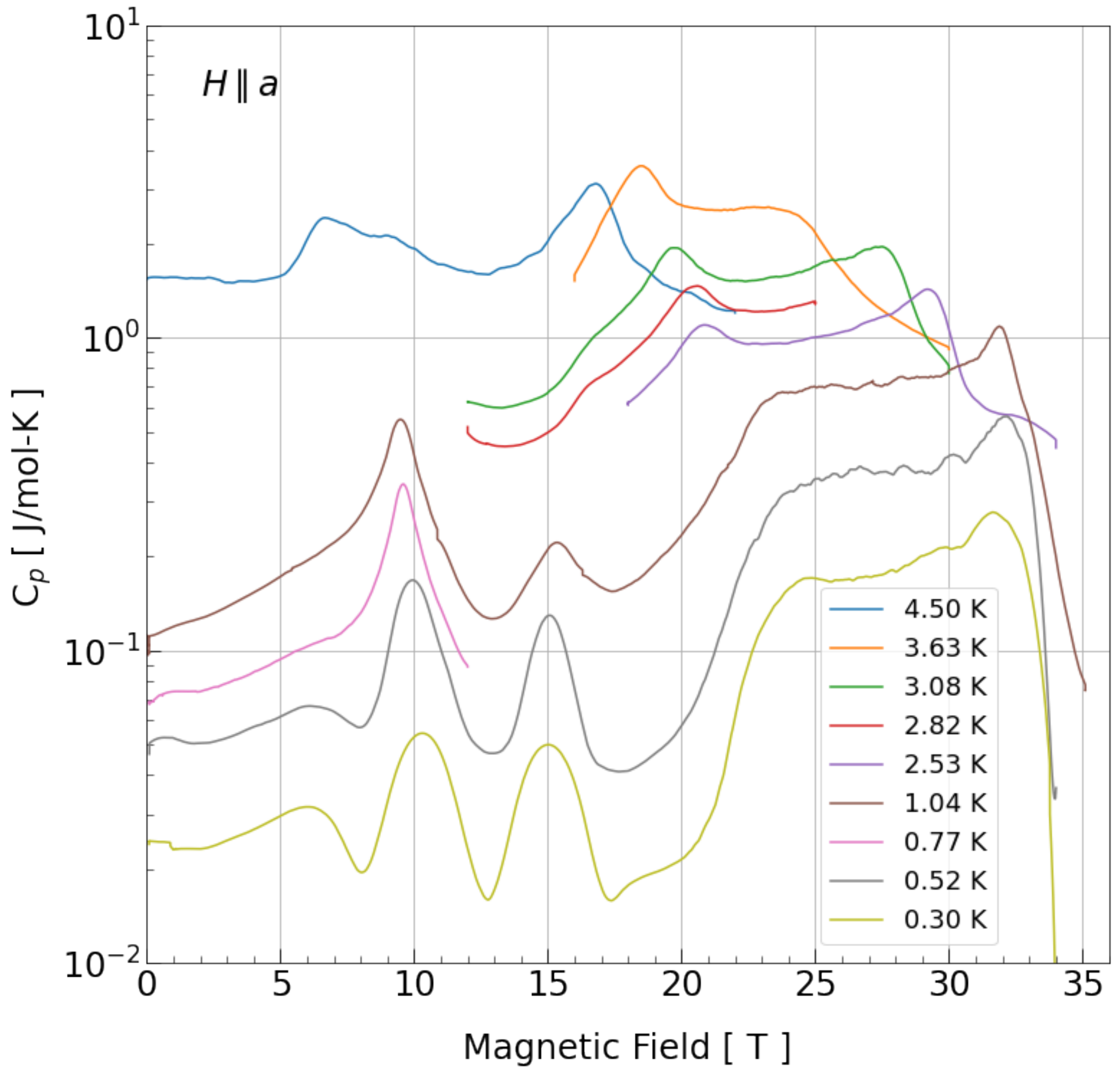}
\caption{(Color online) Magnetic-field dependence of the specific heat as a function of temperature for $H||a$. Features corresponding to magnetic-field induced phase transitions are observed at 6, 10, 15, 24, and 33 T in the low temperature limit. The 6 T transition has not been previously reported.}
\label{fig:Cp_for_H||a}
\end{figure}

The magnetic field dependence of the specific heat at 0.3 K  presented in Fig.~\ref{fig:Cp_for_H||a} for $ H ||a$ ($\phi$ = \ang{90} plane-parallel orientation) reveals a series of magnetic-field-induced phase transitions, including a previously unreported transition at 6 T, transitions into and out of the \textit{UUD} phase at 10 T and 15 T, a sharp rise in the specific heat followed by a plateau corresponding to a transition around 24 T, and finally a sharp drop in the specific heat corresponding to a transition out of the antiferromagnetic state at the saturation field of 33 T.  Additional features seen in the 0.3 K field sweep may correspond to emerging transitions but are not resolved at higher temperature. 

\begin{figure}[ht]
\centering\includegraphics[clip,width=8cm]{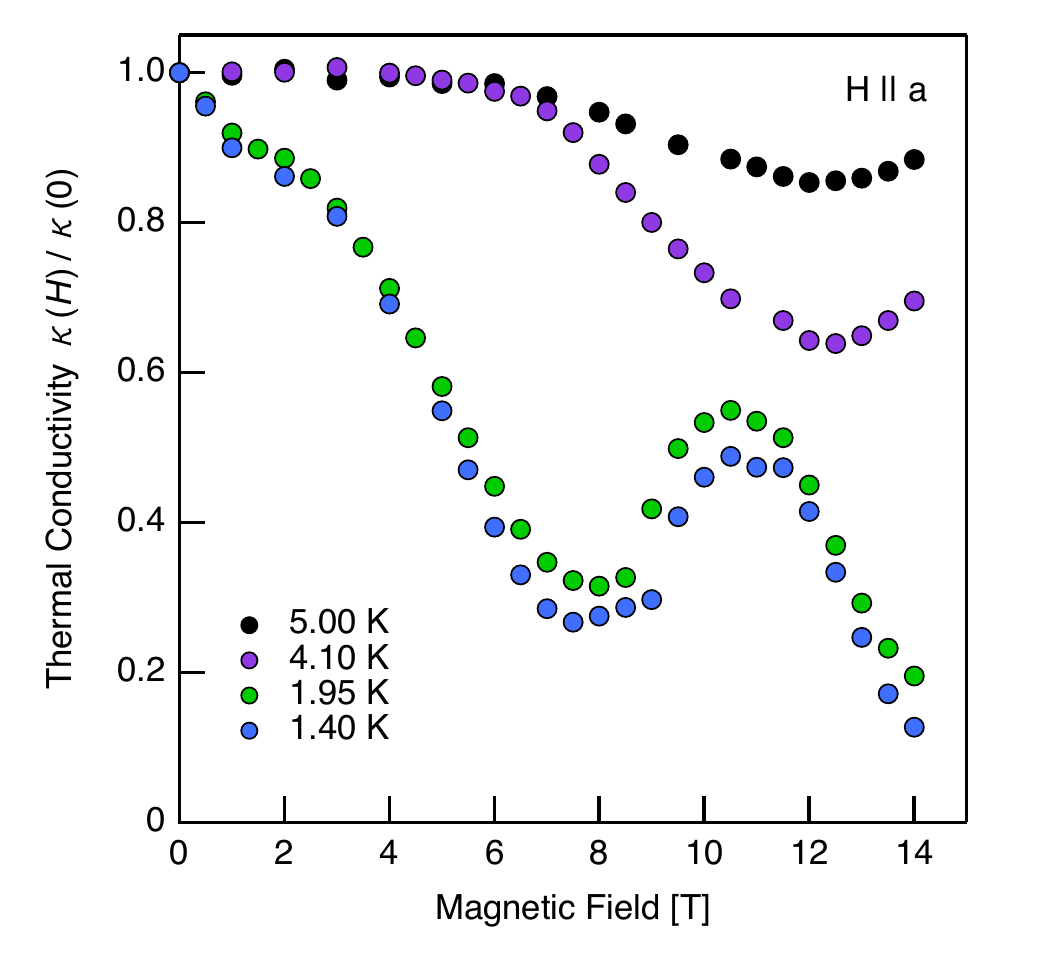}
\caption{(Color online) Magnetic-field dependence of the thermal conductivity for $H||a$ at temperatures above and below the zero-field magnetic ordering temperature $T_N(0) = 3.7$ K, revealing field induced phase transitions from the zero field paramagnetic $T > T_N(0)$ and antiferromagnetic $T<T_N(0)$ states to the \textit{UUD} spin state.   }
\label{fig:K_for_H||a_high_T}
\end{figure} 

The magnetic-field-induced phase transition from the low field orderings to the \textit{UUD} spin ordering can also be seen in thermal conductivity measurements  $\kappa(H)/\kappa(0)$ for $H||a$. Figure~\ref{fig:K_for_H||a_high_T} shows the field dependence for temperatures above 1 K.  At temperatures above $ T_N(0) = 3.7$ K, an applied magnetic field induces a phase transition directly from the lower field paramagnetic state into the \textit{UUD} spin state (at approximately 7 T for the 4.1 K trace); the initial field dependence  below the zero-field ordering temperature $T_N(0)$ arises from change in spin orientation within the low field antiferromagnetic state with applied field. The peak in the thermal conductivity at still higher magnetic field corresponds to the transition from this low field state into the \textit{UUD} spin state  (at approximately 9 T for the 1.95 K trace), in good agreement with the specific heat results presented earlier. 

\begin{figure}[h!]
\centering\includegraphics[clip,width=8cm]{{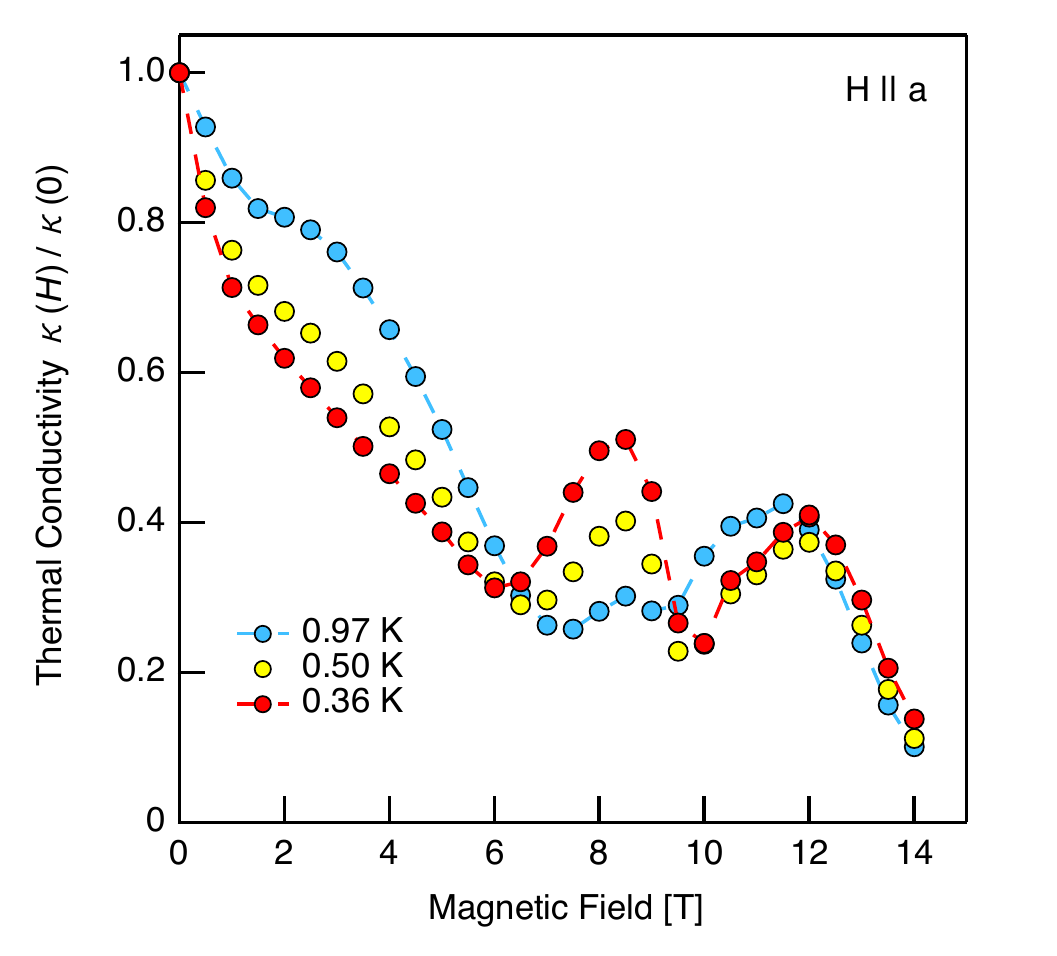}}
\caption{(Color online) Magnetic field dependence of the thermal conductivity for $H||a$ revealing the onset of a new magnetic phase between 6 and 10 tesla below 1 K. }
\label{fig:K_for_H||a_low_T}
\end{figure}

Below 1 K, two peaks are seen as a function of magnetic field, as shown in Fig.~\ref{fig:K_for_H||a_low_T}. The first of these occurs between 6 T and 10 T; the second of these begins at 10 T --- the onset of the \textit{UUD} phase --- and continues up to the maximum measured field of 14 T.  These results match what is seen in our specific heat measurements, further supporting our identification of a new low field phase just below \textit{UUD} at temperatures below 1 K.  

\begin{figure}[h!]
\centering\includegraphics[clip,width=8cm]{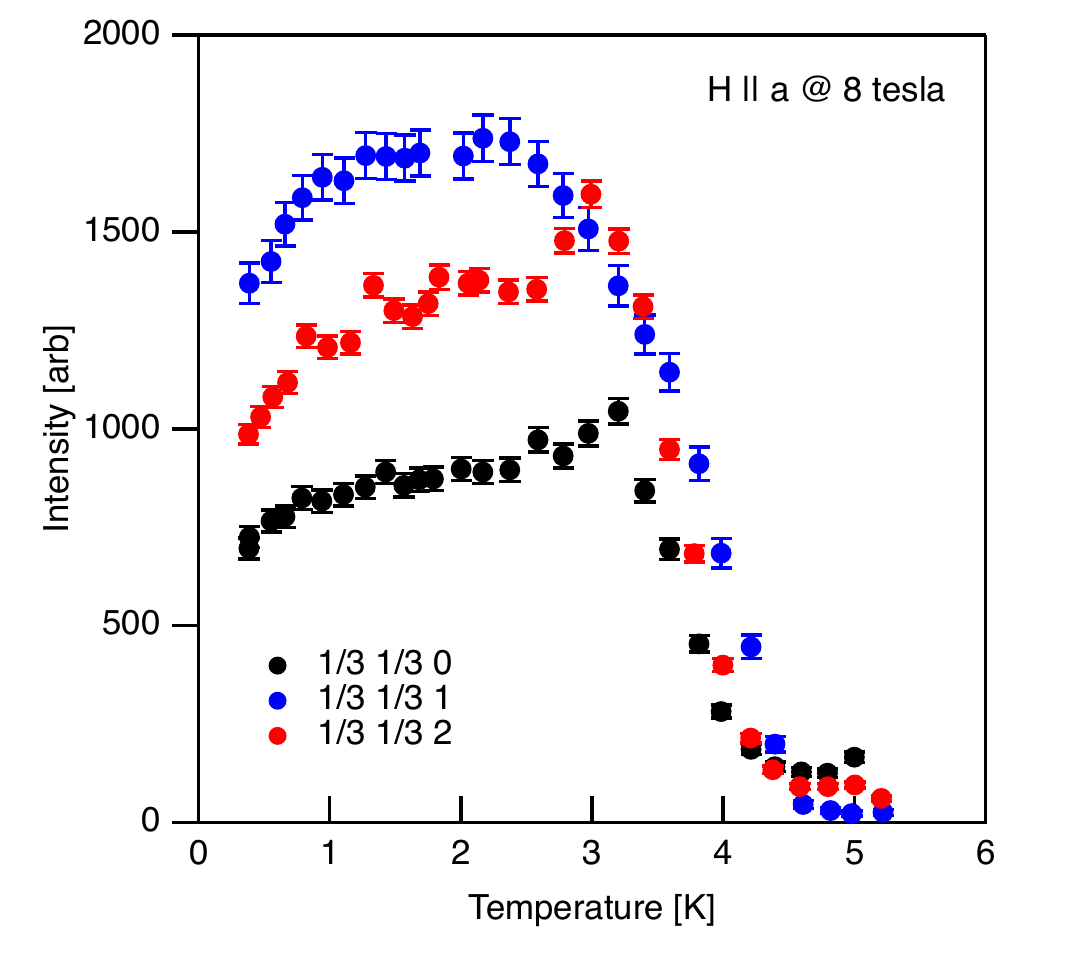}
\caption{(Color online) Intensities of three magnetic Bragg scattering peaks — (1/3, 1/3, 0), (1/3, 1/3, 1), and (1/3, 1/3, 2) — at 8 T as a function of temperature for $H||a$.}
\label{fig:neutron_8T_Bragg}
\end{figure}

 Neutron scattering measurements provide additional information regarding the low field spin orderings.  Temperature dependent measurements at 8 T shown in Fig.~\ref{fig:neutron_8T_Bragg} reveal the onset of antiferromagnetic order in field at temperatures below 4.5 K followed by a pronounced decrease in  interplanar scattering intensity at temperatures below 1 K. The \textit{L} dependence of this decrease --- much steeper for the (1/3, 1/3, 1) and (1/3, 1/3, 2) magnetic Bragg scattering peaks than for the in-plane (1/3, 1/3, 0) scattering peak —  suggests a change in the periodicity of the spin order along the $c$ axis at 1 K. 

In contrast, NMR measurements as a function of magnetic field see only a second-order transition from a low-field `distorted combined \textit{Y}'  phase (Fig.~\ref{fig:2D_spin_with_interlayer}b) to the \textit{UUD} phase \cite{koutroulakis2015quantum} (Fig.~\ref{fig:2D_spin_with_interlayer}e). Those measurements, however, were done at 1.6 K, well above the 1 K transition temperature for this newly discovered phase.  

The possibility of a magnetic-field induced phase transition between two $Y$ phases with differing interplanar coupling below $UUD$ was predicted in an early study  \cite{gekht1997JETP}  but the expected temperature dependence appears inconsistent with what is observed here experimentally. We leave the determination of the true microscopic spin ordering in this low field phase to future experiments. 

\begin{figure}[hb]
\centering\includegraphics[clip,width=8cm]{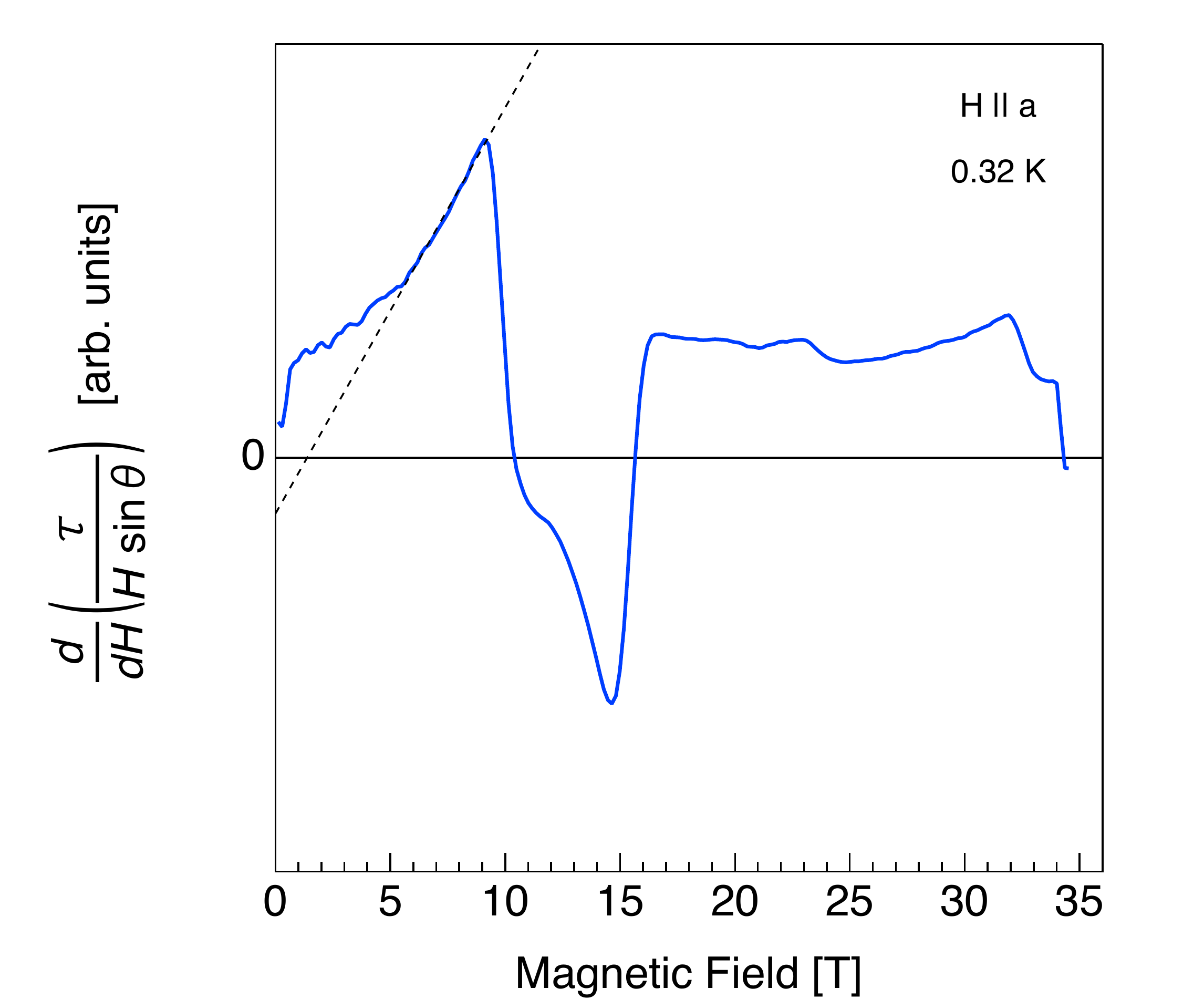}
\caption{(Color online) Magnetic-field dependence of the magnetic torque $\tau$ at 0.32 K for a  $\text{Ba}_3\text{Co}\text{Sb}_2\text{O}_9$ sample for $H || a$ (after scaling for small changes in field orientation). A derivative of $\tau$ has been taken with respect to magnetic field to remove a background term in the magnetometer response.  Phase transitions observed in the specific heat at 6, 10, 15, 24, and 32 T  appear here as extrema and breaks in slope, confirming the magnetic nature of these transitions. The dashed line is a guide to the eye. }
\label{fig:torque||a}
\end{figure}

We now turn to the higher field phase transitions and the construction of a phase diagram for $H||a$. Magnetic torque $\tau$ measurements between 0 T and 35 T at 0.32 K for $H$ oriented close to the $a$ axis provide a complement to specific heat measurements over this same field range. As shown in Fig.~\ref{fig:torque||a}, taking a first derivative of the magnetic torque to remove a background term in the measured response reveals features at fields corresponding to each of the phase transitions observed in the specific-heat measurement, including the new transition observed at 6 T.    We focus here, however, on the transition seen here at 24 T. Calculations assuming alternating-layer six-sublattice spin structures predict, and perhaps only allow,  a first order phase transition at this field \cite{gekht1997JETP,  koutroulakis2015quantum, yamamoto2015microscopic}.  The second order nature of the transition seen here in the specific heat and magnetic torque, while contrary to prediction,  is nonetheless consistent with earlier evidence from magnetization \cite{susuki2013magnetization}, NMR \cite{koutroulakis2015quantum}, and neutron diffraction measurements \cite{liu2019microscopic}.  

In particular, although sometimes taken in the literature to provide evidence for a first order phase transition, magnetization measurements show a continuous $M(H)$ curve with a peak at 24 T rather than a discontinuity in $M(H)$ \cite{susuki2013magnetization}. Further, changes in neutron scattering observed at this field \cite{liu2019microscopic} are consistent with either (a) a first-order transition  from a distorted $V$ spin structure (Fig.~2\ref{fig:2D_spin_with_interlayer}g) to a $V^{\prime}$ structure \cite{yamamoto2015microscopic,liu2019microscopic} (Fig.~2\ref{fig:2D_spin_with_interlayer}h)  or staggered $V$  \cite{koutroulakis2015quantum} (Fig.~2\ref{fig:2D_spin_with_interlayer}i) or  (b) a second-order transition to the $\Psi$ (fan) spin structure (Fig.~2\ref{fig:2D_spin_with_interlayer}j). Finally, NMR measurements reveal a change in spin ordering corresponding to a phase transition at this field but are unable to identify the order of the transition in this particular case \cite{koutroulakis2015quantum}. The identification of the 24 T transition as a first order phase transition has been on the basis of theory, not experiment.   

\begin{figure}[ht]
\centering\includegraphics[clip,width=8cm]{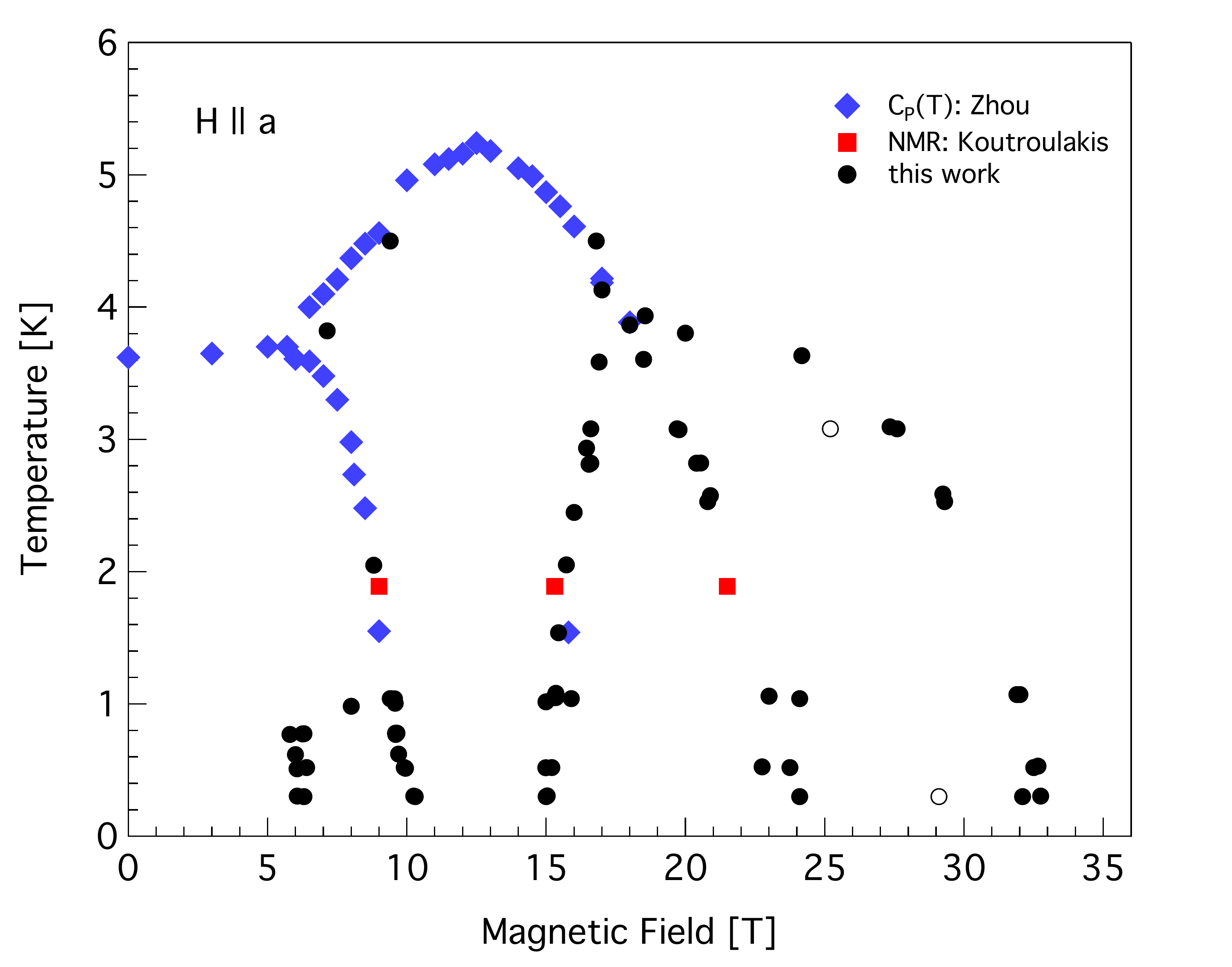}
\caption{(Color online) Phase diagram for $H||a$ ($\phi$ = \ang{90}). Black circles correspond to the locations of phase transitions observed in this work, blue diamonds to earlier, lower field specific heat measurement by Zhou \textit{et al.} \cite{zhou2012successive}, and red squares to NMR measurements by Koutroulakis \textit{et al.} \cite{koutroulakis2015quantum} Hollow black circles correspond to a possible additional high field transition more clearly resolved for $\phi <\ang{90}$. } 
\label{fig:H||a_phase_diagram}
\end{figure}

A magnetic phase diagram for $H||a$  constructed from the results presented above is shown in Fig.~\ref{fig:H||a_phase_diagram}.  In this diagram we also mark the location of a possible additional phase transition at 29 T but stress that this identification is tentative and at the limit of our resolution for $H||a$.

Several features emerge in this phase diagram, most notably the existence of a tetracritical point (at 17 T, 3.8 K) corresponding to the upper vertex of the phase directly above the \textit{UUD} phase in field. In Landau mean-field theory \cite{landau1937JETP, liu1972tetracritical, bruce1975tetracritical, chaikin1995principles}, bicritical and tetracritical points can be found in a number of antiferromagnets in which the lattice anisotropy is not strong enough to produce complete spin alignment. Common examples include spin-flop transitions in which either (a) a single first-order transition line terminating in a bicritical point separates a lower-field antiferromagnetic phase and a spin flop phase (as seen in $\text{Mn}{\text{F}}_2$ \cite{shapira1970MnF2}) or (b) a pair of second-order phase transitions terminating in a tetracritical point bound an intermediate phase that separates a low field antiferromagnetic phase and a high field spin-flop phase (as seen in $\text{Gd}\text{Al}{\text{O}}_3$ \cite{rohrer1977bicritical}). 

A third example -- particularly relevant here -- is when the weakness of the interlayer coupling in a layered structure quasi-2D antiferromagnet with in-plane isotropy leads to a high-field tetracritical point in the magnetic phase diagram  (as previously seen in $\text{K}_2\text{Mn}_{0.978}\text{Fe}_{0.022}\text{F}_4 $ \cite{bevaart1979neutron}). In layered structure triangular lattice quantum Heisenberg antiferromagnets such as $\text{Ba}_3\text{Co}\text{Sb}_2\text{O}_9$, this weak interlayer coupling can lead to a field-induced phase transition that doubles the period of the magnetic structure along the direction normal to the layer. This happens, for example, when the spins alternate directions in alternating layers  \cite{gekht1997JETP,starykh2015unusual}. If second order, the boundary between the bounded intermediate phase and the high field phase and the boundary between the bounded intermediate phase and the \textit{UUD} phase can terminate in a tetracritical point.

Magnetic phases terminating in a tetracritical point indicate breaking of the rotational symmetry of the ideal Heisenberg model, and a symmetry-based argument dictates that the magnetic ordering in the bounded intermediate  phase is related to the ordering in the two  adjacent phases \cite{landau1937JETP, starykh2015unusual}. In determining the nature of the high field spin states, it will therefore be convenient to describe each state in terms of an \textit{order parameter manifold} \cite{starykh2015unusual} representing the symmetries broken by that  spin arrangement. 


The three broken symmetries relevant here are (1) the discrete $Z_3$ symmetry (corresponding, for example, to the choice of sublattice for the down spin in the \textit{UUD} state), (2) the discrete $Z_2$ chiral symmetry  associated with clockwise or counterclockwise rotation, as in the classical cone state, and  (3) the  continuous spin-rotation symmetry $U(1)$  about the axis parallel to the magnetic field. Since commensurate coplanar  states break both $U(1)$ symmetry and $Z_3$ symmetry, the order parameter manifold for both  the \textit{Y} and \textit{V} spin structures is  $U(1)\times Z_3$; the corresponding order parameter manifold for the classical cone (umbrella) state is $U(1) \times Z_2$. In contrast, the commensurate collinear spin structure in the \textit{UUD} phase breaks $Z_3$ symmetry but not $U(1)$ symmetry, so  the order parameter manifold for the \textit{UUD} phase is $Z_3$. 

We can assign an order parameter manifold for the phase directly above the \textit{UUD} phase from NMR \cite{koutroulakis2015quantum} and neutron diffraction \cite{liu2019microscopic}. These measurements indicate that the spin structure in this phase is coplanar and commensurate,  with order parameter $U(1)\times Z_3$; in particular, NMR results are consistent with the distorted \textit{V} \cite{koutroulakis2015quantum} spin structure shown in Fig.~\ref{fig:2D_spin_with_interlayer}g. As a result, the still higher field phase above 24 T can be neither a  collinear phase,  which is ruled out in the first place by the absence of a magnetization plateau in this field region, nor an incommensurate non-coplanar phase, but must instead be either a commensurate coplanar phase (with order parameter  $U(1)\times Z_3$) or an incommensurate coplanar phase (with order parameter $U(1)\times U(1)$).

   Models incorporating interplane exchange and intraplane exchange anisotropy on high field magnetic ordering \cite{ yamamoto2015microscopic, koutroulakis2015quantum} have focused on  the first of these options for the high-field phase: a commensurate coplanar ordering   with order parameter $U(1)\times Z_3$. In particular, for $H||a$, most calculations predict this spin ordering  \cite{yamamoto2015microscopic, koutroulakis2015quantum} just below the saturation field $H_s$ (see however Ref. \cite{gekht1997JETP}). 
 
  The specific type of spin ordering is  dependent on the interplane coupling $J^{\prime}$ and easy-plane anisotropy $\Delta = 1 - J_z / J$, where $\Delta = 0$ for an isotropic Heisenberg antiferromagnet  and $\Delta = 1$ in the XY limit. For easy-plane anisotropies below a critical threshold ${\Delta}_{c1} $ the  previously proposed $V^{\prime}$ spin structure is expected whereas for stronger anisotropies, the coplanar $\Psi$ spin structure occurs. In the absence of interplane exchange $J^{\prime}$,  analysis of a triangular-lattice antiferromagnet with exchange anisotropy near saturation \cite{starykh2014nearsaturation}  predicts that ${\Delta}_{c1} = 0.45/S $   --- corresponding to a transition from  $V^{\prime}$ to $\Psi$ at $ J_z = 0.1 J$ for $S = 1/2$ --- but  including even a small non-zero interplance exchange term $J^{\prime}$ enables this transition to occur for a much smaller degree of in-plane anisotropy $\Delta$ (larger $J_z$), with the  $V^{\prime}$ spin structure favored at lower values of $J^{\prime}$ and the $\Psi$ spin structure at higher values \cite{yamamoto2015microscopic} .   

  An early ESR-based estimate of $J^{\prime}/J \sim 0.03$  \cite{susuki2013magnetization} has been a starting point for most theoretical calculations but more recent neutron scattering experiments at zero field \cite{ma2016static} and in the \textit{UUD} state \cite{kamiya2018nature} consistently establish a higher value of $J^{\prime}/J=0.052$. Similarly, estimates of $J_z/J$ also vary, ranging from  $0.77$ \cite{yamamoto2014quantum, yamamoto2015microscopic} to $0.93$  \cite{susuki2013magnetization}. Within existing theory, the numerical estimates used thus far for $ J^{\prime}$  and $J_z$ have led to the expectation that the high field phase will correspond to the spin arrangement  $V^{\prime}$ rather than  $\Psi$ \cite{koutroulakis2015quantum, yamamoto2015microscopic}. There are, however, a number of unresolved discrepancies \cite{koutroulakis2015quantum} between experiment and theory  for this particular spin order,  
raising the question whether this $V^{\prime}$ identification is  accurate.

  Following \cite{liu2019microscopic}, we therefore now consider both the  $V^{\prime}$ and $\Psi$ arrangements to see if we can choose between the two thermodynamically. A key distinction between the $V^{\prime}$ and $\Psi$ spin structures is this:  in the $V^{\prime}$ type spin arrangements, no spins align directly along the direction of the applied field and the majority spins align at a different angle to the applied magnetic field than  the minority spin, whereas in the symmetric $\Psi$ arrangement, one spin aligns directly with the field while the other two align at equal angles on either side of the first.   The difference in symmetry between the distorted $V$, $V^{\prime}$, and $\Psi$ structures means that the phase transition between the two adjacent phases should be \textit{first order} if between distorted $V$ and  $V^{\prime}$ spin states but \textit{second order} if between distorted $V$ and  $\Psi$  \cite{liu2019microscopic}.  On the basis of the evidence presented here, the spin structure for the phase above 24 T  cannot be the previously theoretically proposed $V^{\prime}$ arrangement but it could be of the  $\Psi$ type. 
 
 As an alternative to the $\Psi$ structure,  we must also  consider  an incommensurate  planar phase with order parameter $U(1)\times U(1)$.  This would not normally be expected in a crystal with sixfold symmetry and the only prior experimental support for this possibility lies in the NMR data,  which although inconclusive,  are  said to bear some resemblance to what would result from incommensurate structures \cite{koutroulakis2015quantum}.  Theoretically, however, commensurate planar to incommensurate planar phase transitions satisfying $U(1)\times U(1)$ have only been predicted for intraplane spatial anisotropy \cite{starykh2014nearsaturation, starykh2015unusual}, which is absent in Ba$_3$CoSb$_2$O$_9$. The experimentally more likely spin ordering for this high field phase is therefore $\Psi$.
 
Distinguishing definitively between a highly symmetric, commensurate planar spin ordering like the  $\Psi$ structure, an incommensurate planar ordering based on one of the commensurate structures proposed for the $V$ phase, or some third yet unconsidered spin structure likely requires new, even higher resolution NMR measurements than reported to date. The broadening of the $^{135}$Ba and $^{137}$Ba spectra with field makes such any measurements challenging, but  recent advances in condensed-matter-NMR compatible high homogeneity high field magnets may make this feasible \cite{bird2015sch}.
 
  \subsection{Results and Discussion for $H||a \rightarrow H||c$}
  
  We turn now to  a discussion of the system's behavior as a function of magnetic field orientation, for orientations from $H||a$ to $H||c$  ($\phi = \ang{90} \rightarrow \phi = \ang{0}$).  Rotating the applied magnetic field out of the easy plane from $H || a $ towards $ H || c$ introduces a transverse field component that can be expected to distort the coplanar and collinear states that occur for $H||a$ and alter the expected locations of the phase boundaries. 

  In the case of the distorted combined \textit{Y} spin structure (Fig.~\ref{fig:2D_spin_with_interlayer}b) experimentally confirmed  by NMR for $H||a$ ($\phi = \ang{90}$) below 10 T at 1.6 K \cite{koutroulakis2015quantum}, it is predicted that the spins will continuously deform with changing $\phi$ into the umbrella phase expected for $H||c$ ($\phi = \ang{0}$). A similar deformation is expected for the \textit{UUD} phase. At higher fields, however, the increasing distortion of  the coplanar phases introduces the possibility of a magnetic-field-orientation-induced phase transition instead of a continuous distortion of the $H||a$ spin structure \cite{koutroulakis2015quantum}. 
  
  \begin{figure}[hbt]
\centering\includegraphics[clip,width=8.0cm]{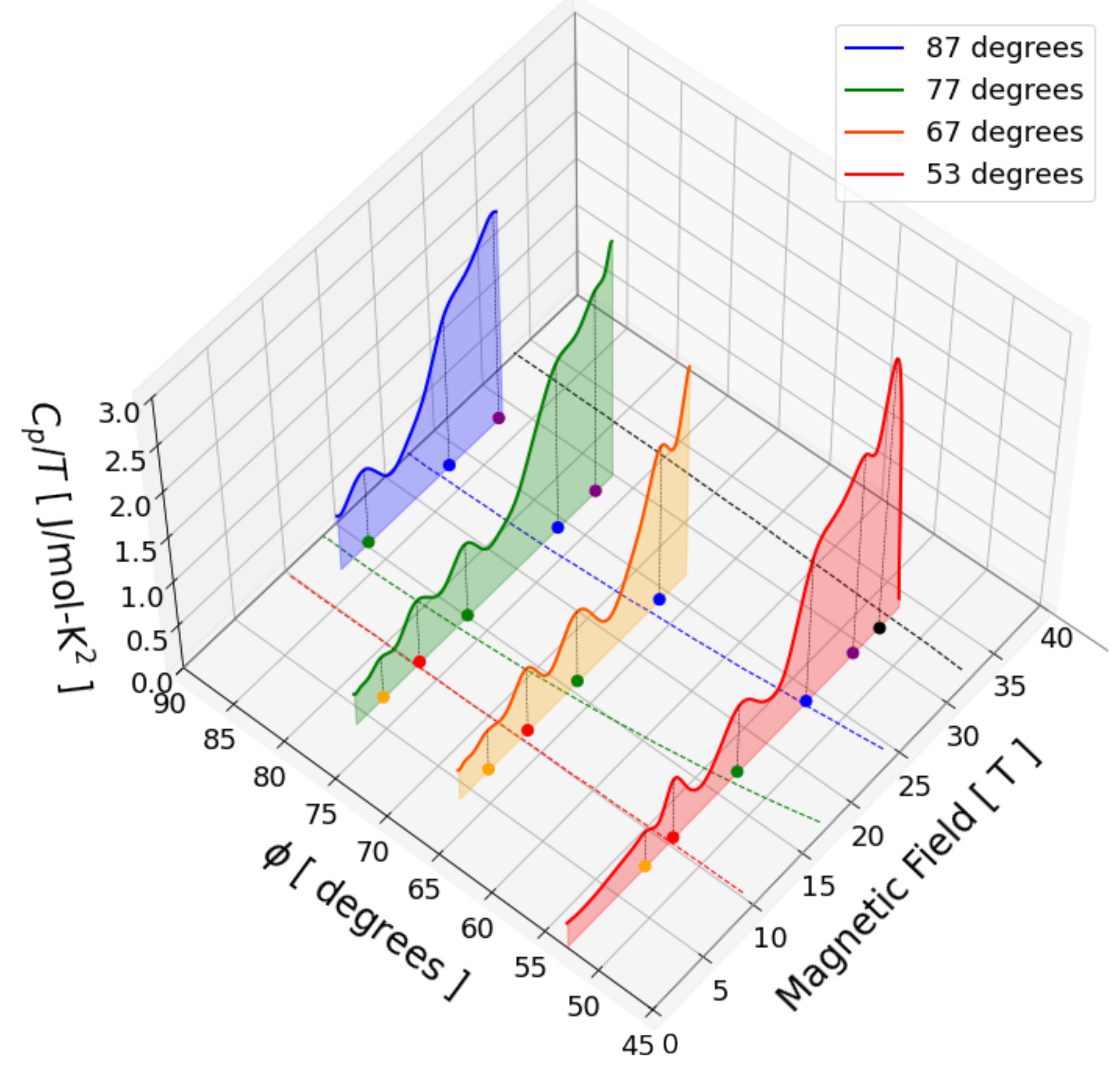}
\caption{(Color online) Variation of the specific heat of $\text{Ba}_3\text{Co}\text{Sb}_2\text{O}_9$, at 0.4 K,  with magnetic field strength and orientation  for magnetic field orientations near $H||a$. The horizontal axes represents magnetic field strength and  magnetic field orientation $\phi$ with respect to the c axis (normal to the easy plane). The horizontal plane serves as an ($H$, $\phi$) phase diagram; the dashed lines on the phase diagram represent the predictions of $T = 0$ semi-classical mean-field theory \cite{koutroulakis2015quantum}. The solid dots on the phase diagram represent locations of experimentally observed magnetic-field-induced phase transitions; dashed vertical lines  leading from the dots mark the corresponding features in the data.}
\label{fig:cp_rotation_near_a}
\end{figure}

\begin{figure}[ht]
\centering\includegraphics[clip,width=8.0cm]{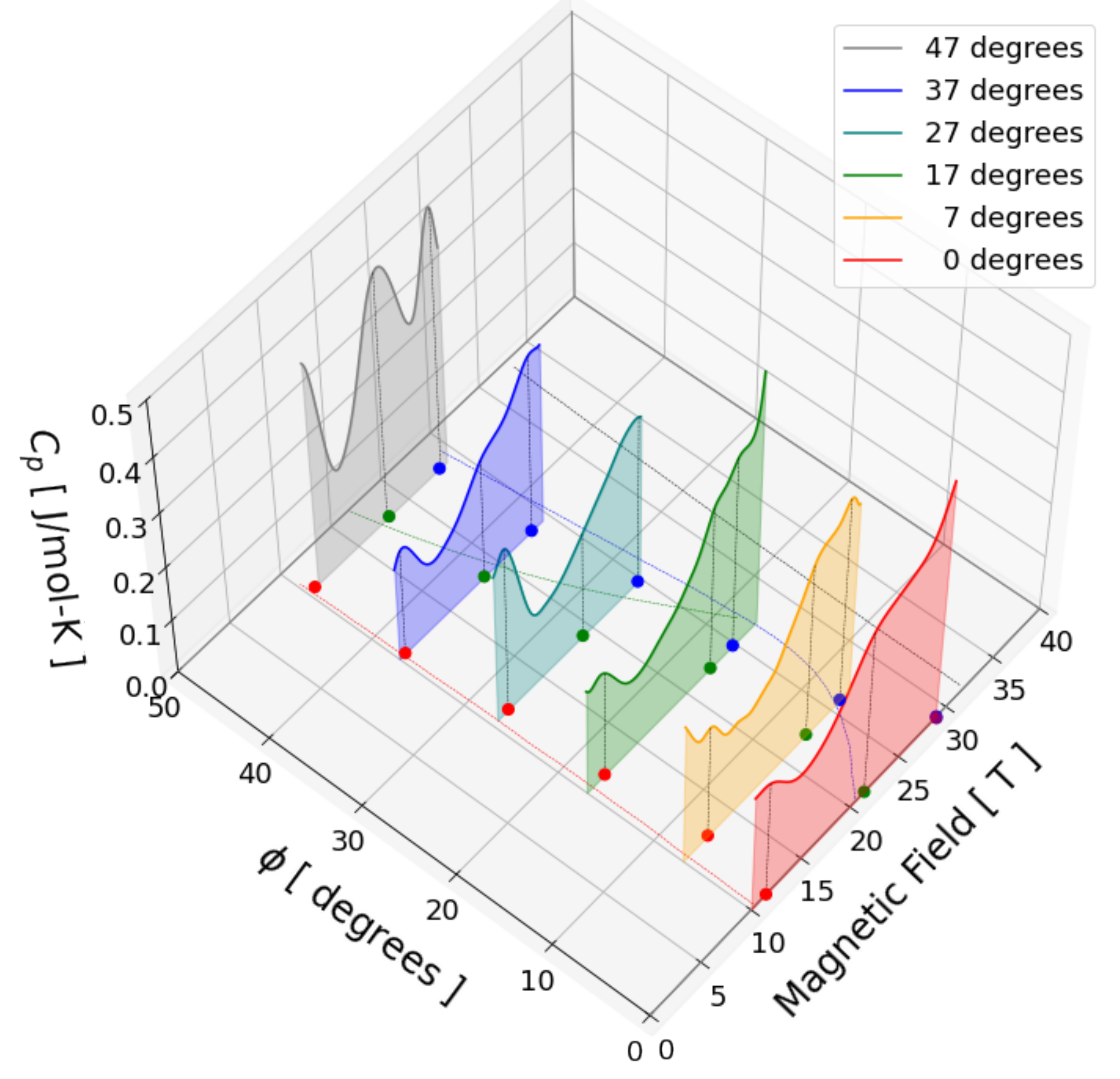}
\caption{(Color online) Variation of the specific heat of $\text{Ba}_3\text{Co}\text{Sb}_2\text{O}_9$   with magnetic field strength and orientation  for magnetic field orientations near $H||c$ ($\phi = 0$) at $T = 1$ K. The horizontal axes represent magnetic field intensity and  magnetic field orientation $\phi$ with respect to the c axis. The horizontal plane serves as an ($H$, $\phi$) phase diagram; the dashed lines on the phase diagram represent the predictions of $T = 0$ semi-classical mean-field theory \cite{koutroulakis2015quantum}. The solid dots on the phase diagram represent locations of experimentally observed magnetic-field-induced phase transitions; dashed vertical lines  leading from the dots mark corresponding features in the data. }
\label{fig:cp_rotation_near_c}
\end{figure}

 Our  results for the angle dependence of the specific heat are shown at 0.4 K in Fig.~\ref{fig:cp_rotation_near_a}  and at 1.0 K in Fig.~\ref{fig:cp_rotation_near_c}. For angles near $H||a$,  the predicted locations in field for the phase boundaries (at $T=0$)  are in good  agreement with our calorimetric measurements shown here, except of course for the new transition  seen at 6 T  and an additional unexpected transition at 29 T.

 In contrast,  we find a  significant deviation between theory and experiment at lower angles: two-layer model calculations of interlayer coupling predict an experimentally unobserved field-rotation-induced phase transition due to the convergence of the lower and upper phase boundaries of an `upper-intermediate-field' (\textit{UIF}) phase — the phase observed between 15 and 24 T for $H||a$ — with decreasing $\phi$. The transition should occur at fields and angles corresponding to the termination of phase boundary shown as a dashed  green line at a dashed blue line in Fig.~\ref{fig:cp_rotation_near_c}. As a result, the \textit{UIF} phase should not exist at all below $\phi_c = \ang{18}$. Experimentally, however, we find that the upper and lower boundaries of the UIF phase fail to converge as a function of $\phi$ before $\phi = 0$.  
 It would be interesting to theoretically investigate how  $\phi_c$ varies with
 a change in interplane coupling parameter $J^{\prime}$ and/or inclusion of next-nearest-neighbor exchange terms.

\subsection{Results and Discussion for $ H || c$}

\begin{figure}[htb]
\centering\includegraphics[clip, width=8 cm]{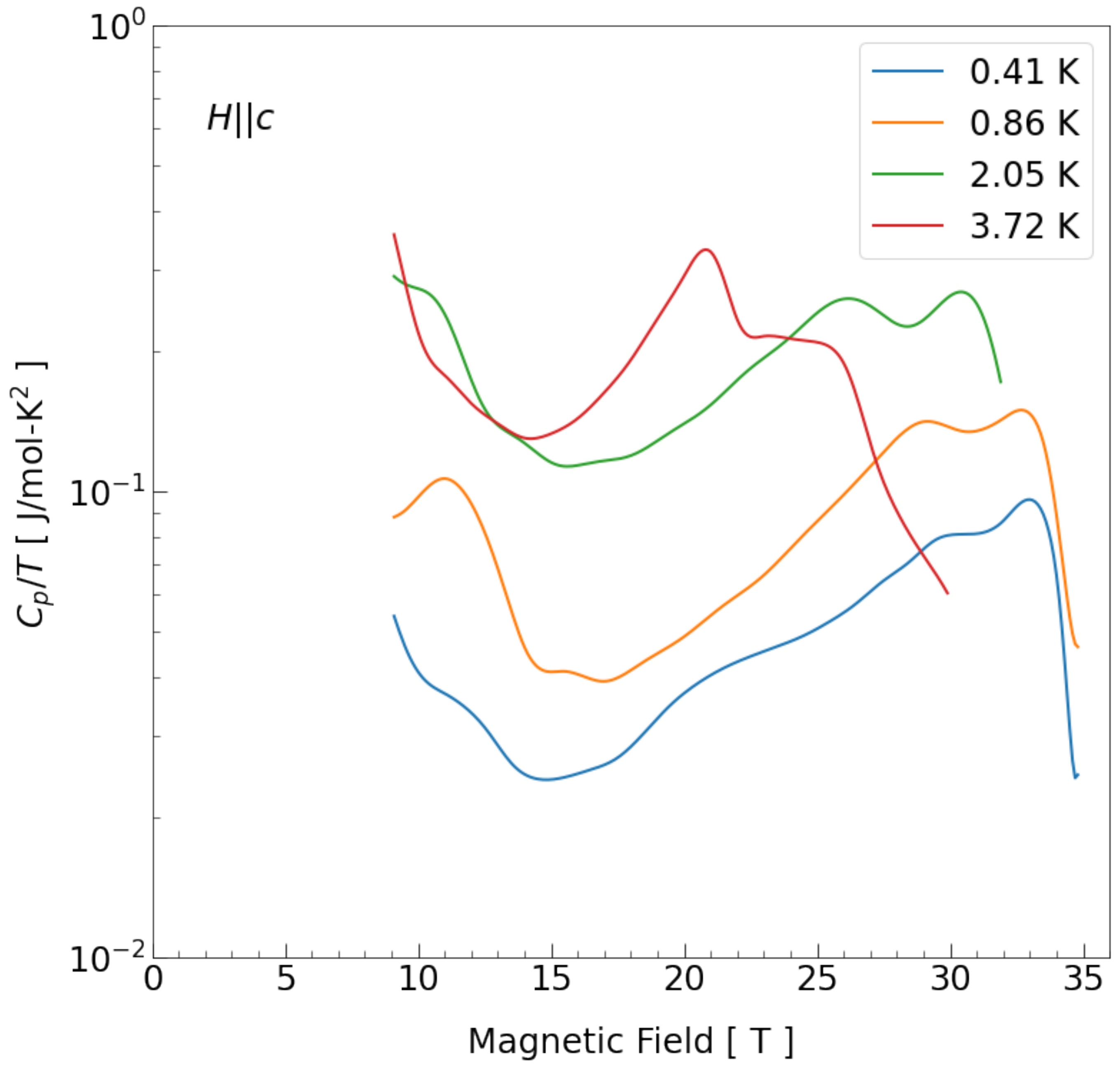}
\caption{(Color online) Magnetic-field dependence of $C_p/T$ for $H||c$ from 9 T --- just below the $UUD$ phase transition at 12 T --- up to the saturation field $H_s = 33$ T. Features corresponding to magnetic-field induced phase transitions are observed at 12, 20, 29, and 33 T at 0.41 K. Lower-field data was not collected for this field orientation. }
\label{fig:specific_heat_H||c}
\end{figure}

Our measurements at $H||c$ are more limited in scope but still provide some additional insights regarding these various phase boundaries.  The magnetic field dependence of the specific heat between 9 and 35 T for temperatures ranging from 0.4 K to 3.7 K --- shown in  Fig.~\ref{fig:specific_heat_H||c} --- reveal magnetic-field-induced phase transitions  at low temperature at 12, 21, 29, and 33 T, the last of these being the transition at the saturation field $H_s$. NMR measurements at 1.89 K were first to reveal the existence of the 21 T transition predicted by theory; the same set of NMR measurements were also the first to raise the possibility of an additional transition somewhere between 27.5 T and 29 T for this field orientation \cite{koutroulakis2015quantum}.

\begin{figure}[hb]
\centering\includegraphics[clip,width=8cm]{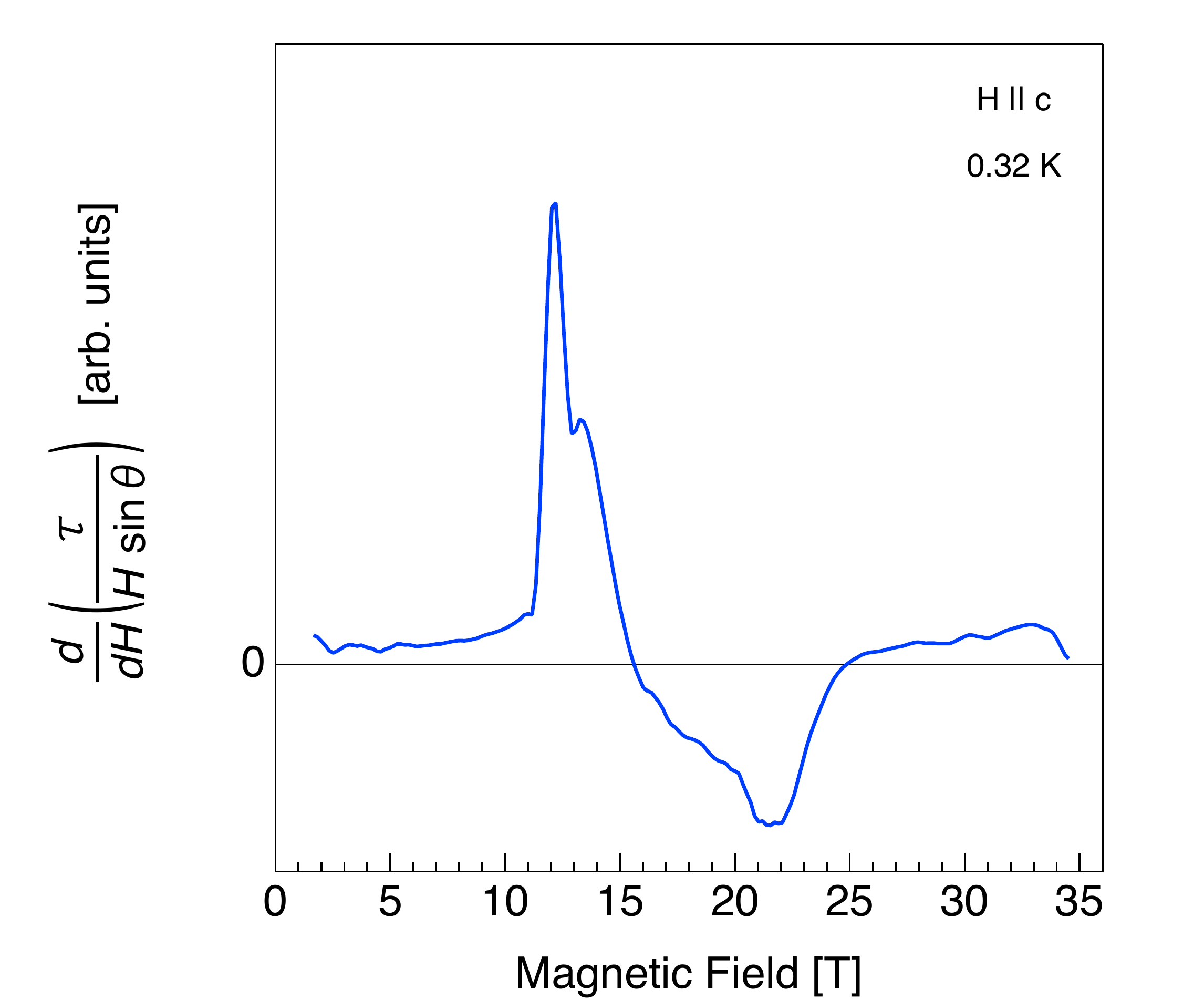}
\caption{(Color online) Magnetic-field dependence of the torque $\tau$ at 0.32 K for $H||c$.  Phase transitions observed in the specific heat for this field orientation at 12, 21, and 33 T are manifest here as extrema and breaks in slope, confirming the magnetic nature of these transitions.}
\label{magnetic_torque_H||c}
\end{figure}

As before, the magnetic torque measurement for $H||c$ at 0.32 K shown in Fig.~\ref{magnetic_torque_H||c} confirms the magnetic character of these transitions. Transitions are observed at 12, 21, and 33 T. The additional transition at 29 T seen in NMR \cite{koutroulakis2015quantum} and also  in the specific heat is too weak to resolve above the noise floor here in magnetization.
 
\begin{figure}[htb]
\centering\includegraphics[clip,width=8 cm]{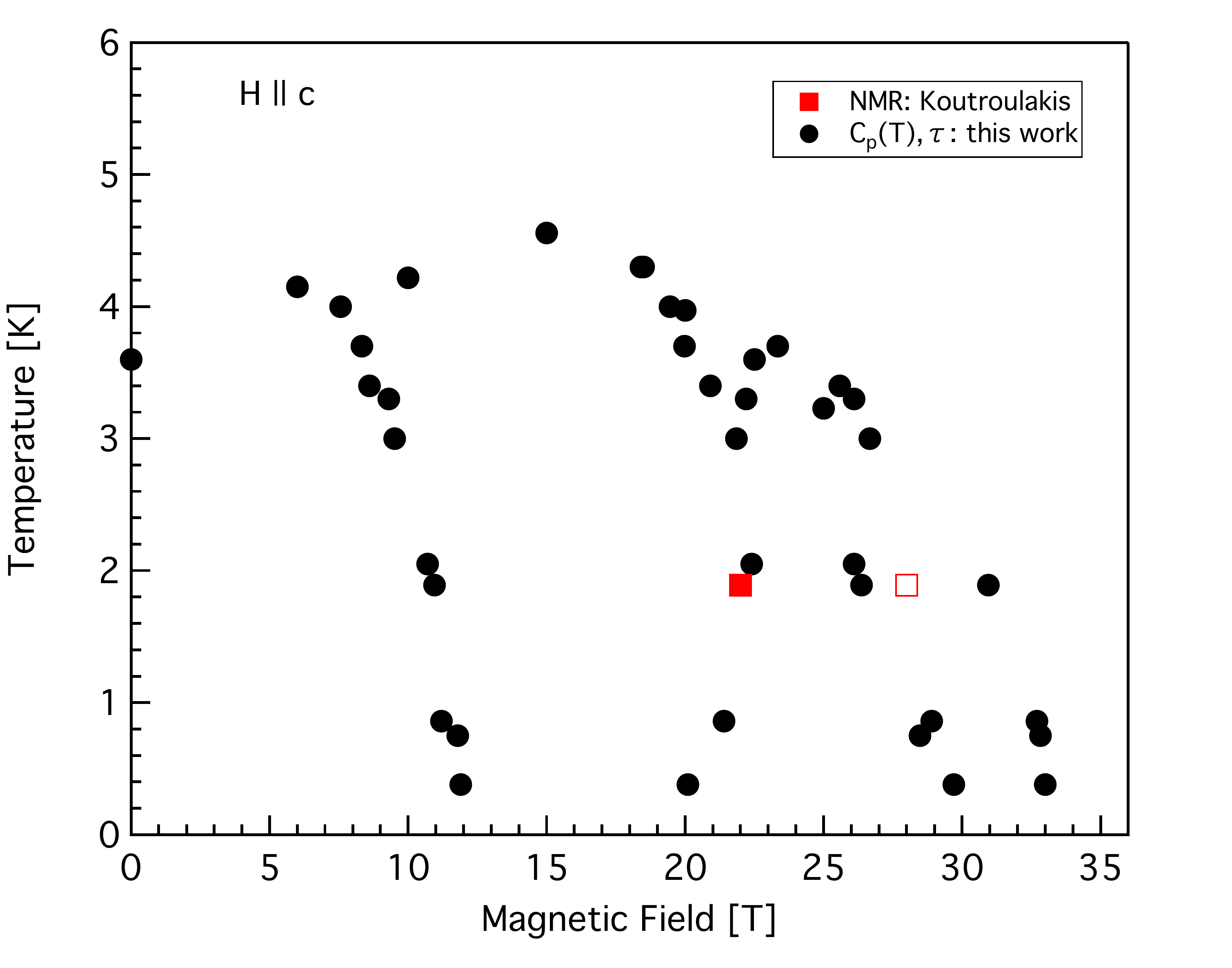}
\caption{(Color online) Phase diagram for $H||c$ ($\phi$ = \ang{0}). Black circles correspond the locations of phase transitions observed in this work, and solid (hollow) red circles represent transitions definitively (tentatively) identified by NMR measurements \cite{koutroulakis2015quantum} at 1.89 K between 14.5 and 30 T.} 
\label{fig:H||c_phase_diagram}
\end{figure}

We present in Fig.~\ref{fig:H||c_phase_diagram} a phase diagram constructed from these measurements. Additional measurements of thermal conductivity up to 14 T (not shown here) are consistent with this phase diagram. We also include two previously reported features in NMR measurements at 1.89 K between 14.5 T and 30 T \cite{koutroulakis2015quantum}.  The NMR measurements are in good agreement with the specific heat and magnetic torque results presented here, although the 29 T transition was identified as tentative \cite{koutroulakis2015quantum}. For $H||c$, the lowest field phase below 12 T was identified as the classically expected umbrella (or cone) spin structure; measurements at the same temperature in the intermediate field phase between 12 and approximately 21 T  are consistent with the coplanar distorted \textit{V} state \cite{koutroulakis2015quantum} shown in Fig.~\ref{fig:2D_spin_with_interlayer}g). Measurements in the high field phase between 21 to approximately 29 T, however, are inconsistent with the theoretically predicted staggered \textit{V} (Fig.~\ref{fig:2D_spin_with_interlayer}i)  spin structure \cite{koutroulakis2015quantum}. The nature of the spin state above 29 T in this orientation has also not been established.  A transition just below $H_s$ to a high field cone phase was predicted in an early work \cite{gekht1997JETP} but has not been reproduced in more recent theoretical calculations \cite{starykh2014nearsaturation, koutroulakis2015quantum, yamamoto2015microscopic}.

\section{general discussion}

Our results indicate that the present $S = 1/2$, two layer $J - J_z -J^{\prime}$ understanding of the triangular lattice quantum Heisenberg antiferromagnetic model captures many aspects of the behavior of $\text{Ba}_3\text{Co}\text{Sb}_2\text{O}_9$ but must now be modified or expanded in some way to account for the second order nature of the phase transitions observed here, the unexpected emergence of additional phase transitions at low temperature, and the evolution of the phase diagram with field orientation. We suggest three possibilities:  relaxing  the assumption of an effective $S = 1/2$ spin system for the Co$^{2+}$ ions,  incorporating next nearest exchange terms, and other more general modifications of the alternating layer six sublattice structure used in calculations so far. 
 
First,  the magnetic properties of the Co$^{2+}$ ions in $\text{Ba}_3\text{Co}\text{Sb}_2\text{O}_9$ are associated with a well-separated Kramers doublet ground state \cite{shirata2012experimental}, leading to the approximation of $\text{Ba}_3\text{Co}\text{Sb}_2\text{O}_9$ as an effective $S= 1/2$ spin system. Including higher-order spin-orbital exchanges, however,  alters the expected magnetic ordering \cite{elliot2020CoTiO3} in a related Co$^{2+}$ easy-plane honeycomb quantum magnet with significant interplane coupling: ${\text{CoTiO}}_3$. Could including higher-order spin-orbital exchanges alter the expected magnetic ordering in $\text{Ba}_3\text{Co}\text{Sb}_2\text{O}_9$ as well? 

Second,  including a next-nearest intraplane exchange term $J_2$ as small as  $J_2/J \simeq 0.07$ is known to destroy the \ang{120} zero field magnetic ordering otherwise expected for the $S= 1/2$ triangular lattice antiferromagnet \cite{hu2019THAF}. It would be interesting to investigate what happens for a still smaller non-zero $J_2$ term which preserves the zero-field ordering \cite{ma2016static}, if a magnetic field is applied. 
 
Third, 
present calculations assume an alternating layer, six-sublattice spin structure for the multilayer triangular lattice quantum Heisenberg antiferromagnet model.
Modifying this 
could allow
periodicities in the spin structures other than simple aligned/anti-aligned order, including the possibility of incommensurate order \cite{kenzelmann2007RbFe(MoO4)2}. Since multiple periodicities can have the same exchange energy \cite{svistov2006RbFe(MoO4)2}, the periodicity could then be selected by some other interaction weaker than nearest-neighbor exchange coupling, such as next nearest exchange or dipole-dipole interactions. These have been seen to lead to differences in the expected magnetic-field-induced phase transitions \cite{svistov2006RbFe(MoO4)2}, an example being the $S = 5/2$ triangular lattice compound $\text{RbFe}{(\text{Mo}{\text{O}}_4)_2}$. In this material, as with $\text{Ba}_3\text{Co}\text{Sb}_2\text{O}_9$,  two phases 
appear below the \textit{UUD} phase for $H||a$ but only one --- the umbrella phase --- for $H||c$ \cite{svistov2006RbFe(MoO4)2, smirnov2007RbFe(MoO4)2, smirnov2017RbFe(MoO4)2}.  

Further experimental work is also needed: the pre-pandemic experiments presented above necessarily focused on mapping out the angle dependence of the previously theoretically expected phase boundaries in the limited magnet time available, leading us to constrain the range of most field sweeps for $\phi \neq \ang{0}$ to fields above 9 T. Post-pandemic, we hope to carry out systematic investigations of the 29 T feature seen near saturation field for some field angles as well as the low field umbrella phase that exists for $H||c$ below 12 T  \cite{koutroulakis2015quantum}, including the interesting question whether the low field 6 T transition seen for $H||a$ persists as $H \rightarrow c$. 

%

\begin{acknowledgements}

Q.H. and H.Z. thank the support from the National Science Foundation through award DMR-2003117. Y.T. thanks the NHMFL UCGP for support. X.F.S. thanks the support from the National Natural Science Foundation of China (Grants No. U1832209 and No. 11874336), the National Basic Research Program of China (Grant No. 2016YFA0300103) and the Innovative Program of Hefei Science Center CAS (Grant No. 2019HSC-CIP001). J.M. thanks the support from the National Natural Science Foundation of China (Grants No. 11774223, No. U1732154, and No.U2032213), the National Basic Research Program of China (Grants No. 2016YFA0300501 and No. 2018YFA0704300) and a Shanghai talent program. A portion of this work was performed at the National High Magnetic Field Laboratory, which is supported by the National Science Foundation Cooperative Agreement No. DMR-1644779 and the State of Florida. A portion of this research used resources at the High Flux Isotope Reactor, a DOE Office of Science User Facility operated by the Oak Ridge National Laboratory.

\end{acknowledgements}

\bibliography{Ba3CoSb2O9}

\end{document}